\documentclass[reprint,
amsmath,amssymb,
aps,
floatfix,
superscriptaddress,
]{revtex4-2}

\usepackage[export]{adjustbox}[2011/08/13]
\usepackage{graphicx} 
\usepackage{dcolumn} 
\usepackage{colordvi}
\usepackage{color}
\usepackage{pstricks}
\usepackage{epstopdf}
\usepackage{amssymb}
\usepackage{url}
\graphicspath{{ps}}
\usepackage{hyperref}
\usepackage{tabularx}
\usepackage{multirow}
\usepackage{siunitx}
\usepackage{hyphenat}
\usepackage{subfigure}
\usepackage{graphicx,float}
\usepackage[italic]{hepnames}
\usepackage{color, colortbl}
\usepackage{mathtools}
\usepackage{orcidlink}
\usepackage[normalem]{ulem}

\newcommand{\Bpipi}{\ensuremath{\overline{B}{}^0 \to \pi^0 \pi^0}\xspace}

\input ./belle2sym.tex

\begin{document}

\def\belletwo {\it {Belle II}}

\vspace*{-3\baselineskip}





\title {Measurement of the branching fraction and $\it CP$-violating asymmetry of the decay $B^{0} \rightarrow \pi^{0} \pi^{0}$ using $387$ million bottom-antibottom meson pairs in Belle II data}

  \author{I.~Adachi\,\orcidlink{0000-0003-2287-0173}} 
  \author{L.~Aggarwal\,\orcidlink{0000-0002-0909-7537}} 
  \author{H.~Ahmed\,\orcidlink{0000-0003-3976-7498}} 
  \author{H.~Aihara\,\orcidlink{0000-0002-1907-5964}} 
  \author{M.~Alhakami\,\orcidlink{0000-0002-2234-8628}} 
  \author{A.~Aloisio\,\orcidlink{0000-0002-3883-6693}} 
  \author{N.~Althubiti\,\orcidlink{0000-0003-1513-0409}} 
  \author{N.~Anh~Ky\,\orcidlink{0000-0003-0471-197X}} 
  \author{D.~M.~Asner\,\orcidlink{0000-0002-1586-5790}} 
  \author{H.~Atmacan\,\orcidlink{0000-0003-2435-501X}} 
% ...
\collaboration{The Belle II Collaboration}

\begin{abstract}
\vspace{0.4cm}
We measure the branching fraction and $\it CP$-violating flavor-dependent rate asymmetry of $B^{0} \to \pi^{0} \pi^{0}$ decays reconstructed using the Belle II detector in an electron-positron collision sample containing $387 \times 10^{6}$ $B\overline{B}$ pairs. Using an optimized event selection, we find $126\pm 20$ signal decays in a fit to background-discriminating and flavor-sensitive distributions. The resulting  branching fraction is $(1.25 \pm 0.23)\times 10^{-6}$ and the $\it CP$-violating asymmetry is $0.03 \pm 0.30$.

\vspace{0.5cm}
\end{abstract}


\maketitle

{\renewcommand{\thefootnote}{\fnsymbol{footnote}}}
\setcounter{footnote}{0}


The decay of the neutral bottom-meson into a pair of neutral pions, $B^0 \rightarrow \pi^0 \pi^0$~\footnote{Charge-conjugate modes are implicitly included unless noted otherwise throughout this paper.}, plays an important role in the study of the flavor-changing weak interactions of quarks. The decay properties can be used to test and refine phenomenological models of hadronic bottom-meson amplitudes and to provide important inputs for the determination of $\phi_2$, a relevant parameter in weak quark dynamics. 
These measurements are expected to significantly impact current constraints on potential processes not described by the standard model (SM). 

Theoretical predictions for the branching fraction $\mbox{\ensuremath{\mathcal{B}(B^0 \rightarrow \pi^0 \pi^0)}}$ are challenging because the calculation of hadronic amplitudes involves low-energy, nonperturbative gluon exchanges. Currently available approximate methods often fail to reproduce data. Predictions based on QCD factorization~\cite{Beneke:1999br,Beneke:2003zv,Beneke:2005vv,Pilipp:2007mg} and perturbative QCD~\cite{Lu:2000em, Zhang:2014bsa} are approximately five times smaller than the experimental results. In addition, the ratio of color-suppressed to color-allowed tree amplitudes, as inferred from other charmless two-body decay modes, does not agree with expectations~\cite{Charng:2004ed}, possibly indicating anomalously large contributions~\cite{Mishima:2004um, Charng:2003iy}. An improved understanding of the $B^0 \to \pi^0\pi^0$ decay amplitudes could be relevant to the so-called $\B \to K\pi$ puzzle~\cite{Li:2005kt,Baek:2009pa,Barger:2004hn}.

The study of $\it CP$-violating asymmetries in the rates of two-body decays involving the $\mbox{\ensuremath{b \to u}}$ transition, such as $B^0 \to \pi^0\pi^0$, is also relevant. These asymmetries currently offer reliable and precise access to the least-well-known angle of the unitarity triangle, $\phi_2 \equiv \textrm{arg}\left( -V_{td} V^*_{tb} / V_{ud} V^*_{ub} \right)$\footnote{The angle $\phi_2$ is also referred to as $\alpha$.}, where $V_{ij}$ are elements of the Cabibbo-Kobayashi-Maskawa (CKM) quark-mixing matrix~\cite{Kobayashi:1973fv,Cabibbo:1963yz}. Improved measurements of $\phi_2$ increase the power of tests of CKM-matrix unitarity and impose more stringent bounds on possible SM extensions. Both $b \to u$ ($W$ emission, or tree) and $b\to d$ ($W$ exchange, or penguin) transitions contribute to the decay amplitude. The determination of $\phi_2$ requires measurements of the branching fractions and {\it CP} asymmetries of all isospin-related $B \to \pi\pi$ decay modes,  i.e., $B^0 \to \pi^+\pi^-$, $B^0 \to \pi^0\pi^0$, and $B^+ \to \pi^+\pi^0$, or of analogous $B \to \rho \rho$ decays, to separate penguin and tree contributions~\cite{Gronau1990, Charles:2017evz}.

 Currently, the uncertainty in $\phi_2$ is dominated  by the uncertainties in the $B^0 \to \pi^0 \pi^0$ branching fraction and $\it CP$-violating flavor-dependent decay-rate asymmetry,
\begin{equation}
    \Acp(B^0 \to \pi^0\pi^0) = \frac{\Gamma(\overline{B}{}^0 \to \piz \piz)-\Gamma(B^0 \to \piz \piz)}{\Gamma(\overline{B}{}^0 \to \piz \piz)+\Gamma(B^0 \to \piz \piz)},
    \label{eq:acp}
\end{equation}
where $\Gamma$ is the decay width.  The world-average values $\mathcal{B}(B^0 \to \pi^0 \pi^0) = (1.55 \pm 0.17) \times 10^{-6}$ and $\mathcal{A}_{\it CP}(B^0 \to \pi^0 \pi^0) = 0.25 \pm 0.20$~\cite{ParticleDataGroup:2024cfk} combine measurements reported by the BaBar~\cite{PhysRevD.87.052009} and Belle~\cite{Belle:2017lyb} collaborations. 


Improved determinations would bring the $B \to \pi\pi$ bounds on $\phi_2$  closer in precision to those from $B \to \rho \rho$, thus improving knowledge of $\phi_2$. They  may also help discriminate among the  scenarios proposed to achieve a consistent picture of the relevant dynamics~\cite{Qiao2015, Li2011, Li2017, Cheng2015}. 

In this paper, we present a measurement~\cite{myThesis} of the branching fraction and {\it CP} asymmetry for the $\Bpipi$ decay using a 365 fb$^{-1}$ sample of electron-positron collision data. The sample contains $387 \times 10^{6} B\overline{B}$ pairs produced near threshold at the $\Y4S$ resonance from 2019 through 2022. A 42.3\,fb$^{-1}$ sample collected at 60 MeV lower energy and hence not containing any $B\overline{B}$ pairs (off-resonance) is also used for background modeling. The samples are collected with the Belle II detector~\cite{Belle-II:2010dht}, located at the SuperKEKB asymmetric-energy collider~\cite{Akai:2018mbz}.

The principal experimental challenge is the reconstruction of a rare decay in a final state with no charged particles, $B^0 \to \pi^0 (\to \gamma\gamma) \pi^0 (\to \gamma\gamma)$. The reconstruction of photons is based solely on calorimeter information and therefore it is much less precise than the reconstruction of charged particles. In addition, it is affected by energy leakage and beam-induced backgrounds due to beam interactions with the beam pipe or residual gas. 
The analysis is developed using simulated samples and signal-free data control regions. The signal region in data is examined only when all procedures are established. Data are enriched in $B^0\rightarrow\pi^0\pi^0$ events by means of an optimized selection based on two statistical-learning classifiers that suppress backgrounds, which are dominated by light-quark production. The flavor of the neutral $B$ meson, needed for the measurement of the {\it CP}-violating asymmetry, is determined using information associated with the other, nonsignal $B$-meson produced in the $\Upsilon(4S)$ decay, since the $\pi^0\pi^0$ final state is common to $B^0$ and $\overline{B}{}^0$. A multidimensional fit of sample composition determines the signal yield and \textit{CP}-violating asymmetry.  The fit results, combined with acceptance and efficiency corrections obtained from simulation and validated using control data, determine the quantities of interest.  We use $D^{*+} \rightarrow D^0(\rightarrow K^- \pi^+ \pi^0) \pi^+$ decays to validate photon reconstruction.  In addition, we validate the analysis using $B^{+} \to K^+ \pi^0$ decays, which yield a final-state \piz with kinematic properties similar to the signal \piz mesons, and $B^{0} \to \overline{D}{}^0(\to K^+ \pi^- \pi^0) \pi^0$ decays, which, like the signal decay, yield two \piz mesons. These $B$ decay modes are ten times more abundant than our expected signal. 

Belle II consists of several subdetectors arranged in a cylindrical structure around the beam pipe~\cite{Belle-II:2010dht}. The $z$ axis is the symmetry axis of a superconducting solenoid, which generates a 1.5~T uniform field along the beam direction. The positive $z$ direction corresponds approximately to the electron-beam direction, and defines the origin of the polar angle $\theta$. The detector structure is divided into three polar regions in increasing order of $\theta$, the forward endcap, barrel, and backward endcap, which correspond to the polar-angle ranges $[12.4,31.4]^\circ$, $[32.2,128.7]^\circ$, and $[130.7,155.1]^\circ$, respectively. The inner volume contains a two-layer silicon pixel detector surrounded by a four-layer double-sided silicon strip detector and a drift chamber. Of the outer pixel layer, only a 15\% azimuthal sector was installed for the data used in this work. These subdetectors are primarily used to reconstruct charged-particle trajectories, and therefore their origin, momentum, and electric charge. These subdetectors also provide charged-particle identification through measurement of specific ionization. A time-of-propagation counter and an aerogel ring-imaging Cherenkov detector covering the barrel and forward endcap regions, respectively, are used for primary charged-particle identification. In this work, the electromagnetic calorimeter is particularly important. It is a segmented array of 8736 thallium-doped cesium iodide crystals arranged in a quasiprojective geometry toward the interaction point.  They occupy the remaining volume inside the superconducting solenoid and cover about 90\% of the solid angle in the center-of-mass (c.m.) frame. The calorimeter identifies electrons and photons in the range of energies 0.02--10 GeV with fractional resolutions of 7.7\% at 100 MeV or $2.2\%$ at 1 GeV~\cite{Miyabayashi:2020xzp}. Resistive plate chambers and scintillating bars are installed in the flux return of the magnet and identify muons and $K^0_L$ mesons. This detector, along with the calorimeter, also contributes charged-particle-identification information.

We use simulated samples to optimize event selection, select fit models, calculate signal efficiencies, and study sources of background. To study signal, we use $2 \times 10^6$ $\Y4S \to \B^0 \overline{B}{}^0$ decays in which one $B$ meson decays into the $\piz \piz$ final state and the other $B$-meson decay is unbiased. These samples are generated with the  EvtGen~\cite{Lange:2001uf} and PYTHIA~\cite{Sjostrand:2014zea} software packages. To study backgrounds, we use simulated samples at least four times larger than the data sample. These samples consist of $\epem \to \Y4S \to B\overline{B}$ processes generated with EvtGen and PYTHIA along with $e^+e^- \to \tau^+ \tau^-$ and continuum $\epem \to \qqbar$ background, where $q$ denotes a $u$, $d$, $s$, or $c$ quark, generated with the KKMC~\cite{Jadach:1999vf}, PYTHIA, and TAUOLA~\cite{Jadach:1993hs} software packages. 
Beam-induced backgrounds sampled from data are included in the simulation~\cite{LIPTAK2022167168}. The detector response is simulated using the GEANT4 ~\cite{GEANT4:2002zbu} software package.
Simulated and experimental data are processed with the Belle II software~\cite{Kuhr:2018lps, basf2-zenodo}. 

The online event selection of our sample requires events to satisfy criteria based on total energy and neutral-particle multiplicity to preferentially retain hadronic events and has full efficiency for the signal decay mode.

In the offline analysis, we identify photon candidates as calorimeter energy deposits (clusters) larger than 30 MeV that involve more than one crystal to reject calorimeter noise. Since CsI(Tl) scintillation light has a relatively long decay time, 
random photons from  hadronic events combined with overlapping residual energy not related to the relevant collision may be misreconstructed as $\pi^0$ candidates. Hence, the time associated with cluster formation is required to be within 200 ns of the collision time. Multiplicative photon-energy corrections ranging between 0.990 and 1.010, with 0.15\%--0.50\% uncertainties, are derived from control samples in data and used to correct for calorimeter-energy miscalibration. 

Selected photons are paired to form \piz candidates, which are further selected to suppress combinatorial background from low-energy photons. We require the \piz momentum in the laboratory frame be greater than 1.5 \gevc, and that the three-dimensional angle between the momenta of final-state photons in the laboratory frame be less than 0.4 radians. In the $\piz$ rest frame, the absolute value of the cosine of the angle between the photon direction and the boost direction from the laboratory frame is required to be less than 0.98 as misreconstructed $\piz$ mesons tend to peak near 1.00. The diphoton mass is required to be between 0.115~$\gevcc$ and 0.150~$\gevcc$, a range of approximately $-2.5$ and $+2.0$ units of resolution around the known \piz mass~\cite{ParticleDataGroup:2024cfk}. The range is asymmetric to compensate for energy leakage from the calorimeter. 
The high momentum of signal $\pi^0$'s offers a natural suppression of misreconstructed and nonsignal photons. To reduce contributions from these sources further, we use a boosted decision tree~\cite{Keck:2017gsv} called ``Photon-BDT". This decision tree is trained on simulated signal photons against beam-background and misreconstructed photons using nine input observables. In approximate order of decreasing discriminating power, they are the energy detected in the crystal having the highest signal, three observables that describe the energy sharing among crystals~\cite{myThesis}, the photon's momentum transverse to the beam direction, the distance between the cluster and the trajectory of the nearest charged particle, the number of crystals in the cluster, the cluster polar direction, and one observable that describes the fraction of cluster energy detected in the central crystal.
We choose the threshold on the Photon-BDT output that maximizes the yield of signal photons over the square-root of the sum of  misreconstructed- and beam-induced- photon yields as expected from simulation.
In simulated samples, this selection removes 83\% of misreconstructed and beam-induced photons and retains 96\% of signal photons. Studies of $D^{*+} \rightarrow D^0(\rightarrow K^- \pi^+ \pi^0) \pi^+$ and $B^{+} \rightarrow K^+ \pi^0$ decays validate the Photon-BDT classifier's performance by showing that signal efficiency and Photon-BDT output distributions are consistent between simulations and data. 

We improve the momentum resolution of the $\pi^0$ candidates by performing a kinematic fit that constrains the diphoton mass to the known $\pi^0$ mass~\cite{ParticleDataGroup:2024cfk}. Signal candidates are reconstructed by combining two $\piz$ candidates.


We use topological observables that exploit the jet-like nature of \qqbar events and the isotropic distribution of final-state particles from $B\overline{B}$ events to reduce the large continuum background, along with observables specifically sensitive to the presence of the nonsignal $B$ meson in the event.  We train a boosted-decision-tree classifier $C$ to discriminate signal from continuum background by analyzing 29 observables comprising, in approximate order of decreasing discriminating power,  modified Fox-Wolfram moments~\footnote{The Fox-Wolfram moments were introduced in \href{https://doi.org/10.1103/PhysRevLett.41.1581}{G. C. Fox and S. Wolfram, Phys. Rev. Lett. 41, 1581 (1978)}. The modified moments used in this paper are described in \href{https://doi.org/10.1103/PhysRevLett.87.101801}{K. Abe et al. (Belle Collaboration) Phys. Rev. Lett. 87, 101801 (2001) and \href{https://doi.org/10.1016/S0370-2693(01)00626-8}{K. Abe et al. (Belle Collaboration), Phys. Lett. B 511, 151 (2001)}.}}, sphericity-related quantities~\cite{PhysRevD.1.1416}, thrust-related quantities~\cite{Farhi:1977sg}, and energy detected in sets of concentric cones with various opening angles centered around the thrust axis~\cite{Brandt:1964sa}. Observables showing correlations larger than 10\% with those used in the sample-composition fit are excluded. The training uses simulated signal samples and off-resonance data. An optimization identifies the selection on $C$ that minimizes the average variance on the branching fraction and {\it CP} asymmetry as expected from repeating the measurement on simplified simulated samples.  The resulting selection on $C$ rejects 98\% of continuum background while retaining 64\% of signal. For convenient modeling, we apply the properties of the probability integral transform~\cite{fisher} to the classifier obtaining $C_{t}$, which has a Gaussian-like shape that peaks at 2.0 for signal and at 0.0 for background~\cite{PhysRevD.107.112009}. 
In addition, we restrict the sample using two kinematic variables that exploit four-momentum conservation in near-threshold $B\overline{B}$ pair production, and are particularly convenient to isolate signal candidates,
    $\mbc = \sqrt{E_{\rm beam}^2/c^4 - |\vec{p}_{B}|^2/c^2}$ and 
    $\dele = E_{B} -  E_{\rm beam}$,
where $E_{\rm beam}$ is the beam energy and $(E_{B},\vec{p}_{B})$ is the four-momentum of the $B$ candidate, both calculated in the \Y4S rest frame. The $\mbc$ and $\dele$ distributions for signal decays peak at the $B$ mass and close to zero, respectively. 
Candidate $B$ mesons are required to have $\mbc > $ 5.2~\gevcc and $-0.3  < \dele < 0.5~\gev$. 

Multiple signal candidates (typically just two) are reconstructed in 0.3\% of events, in which case we choose one candidate at random. 
Following all selections, 27\% of signal events remain, of which 99\% are correctly reconstructed. Such a low misreconstruction rate is due to the low fraction of signal events in which the accompanying  $B$ meson yields a high-momentum $\pi^0$ candidate.

To measure the {\it CP} asymmetry, the $B$-signal flavor is assigned in each event by determining statistically the flavor of the accompanying neutral $B$ meson (tagging). Tagging information is encoded in the $q=+1$ or $-1$ (for $\bar{b}$ or $b$) flavor of the nonsignal $B$ and in $r = 1-2w$, where $w$ is the probability for wrongly tagging an event. The probability is calibrated on control data by measuring $\BzBzb$ flavor oscillations with $B^0 \to D^{(*)-} \pi^+ $ decays~\cite{GFlaT}. We use a novel algorithm~\cite{GFlaT}, which efficiently uses information from nonsignal charged particles and from their relationships. The algorithm improves the average value of $r^2$, the statistical tagging efficacy, by 18\% with respect to the algorithm used in the previous Belle II measurement~\cite{Belle-II:2021zvj}. Unlike in previous analyses where data were analyzed separately in independent contiguous $qr$ intervals (binned flavor tagging), here the unbinned distribution of $w_t$, obtained using the properties of the probability integral transform of $w$, is used as a fit observable. It has a Gaussian-like shape that simplifies the fit model.

The resulting event sample consists of three components, i.e., signal ($s$), continuum ($c$), and background from nonsignal $B$ decays ($B\overline{B}$). The 1\% fraction of genuine signal events that are misreconstructed is classified as signal. Simulated samples show that 81\% of $\B\overline{B}$ background is from $B^+ \to \rho^+ \pi^0$ decays. 
We determine the branching fraction and {\it CP} asymmetry with a likelihood fit of the unbinned distributions of $\mbc$, $\dele$, $C_{t}$, and $w_t$. The signal distribution peaks in each fit observable. The continuum background produces smooth distributions in $\mbc$ and $\Delta E$, and peaks at zero (positive) values in $C_{t}$ ($w$). The $B\overline{B}$ background peaks at similar values of $\mbc$ and $C_{t}$ as the signal, but its $\dele$ distribution is shifted to negative values due to missing energy associated with the unreconstructed decay products.

\begin{figure*}[th!]
    \makebox[0.245\textwidth]{\includegraphics[width=0.245\textwidth]{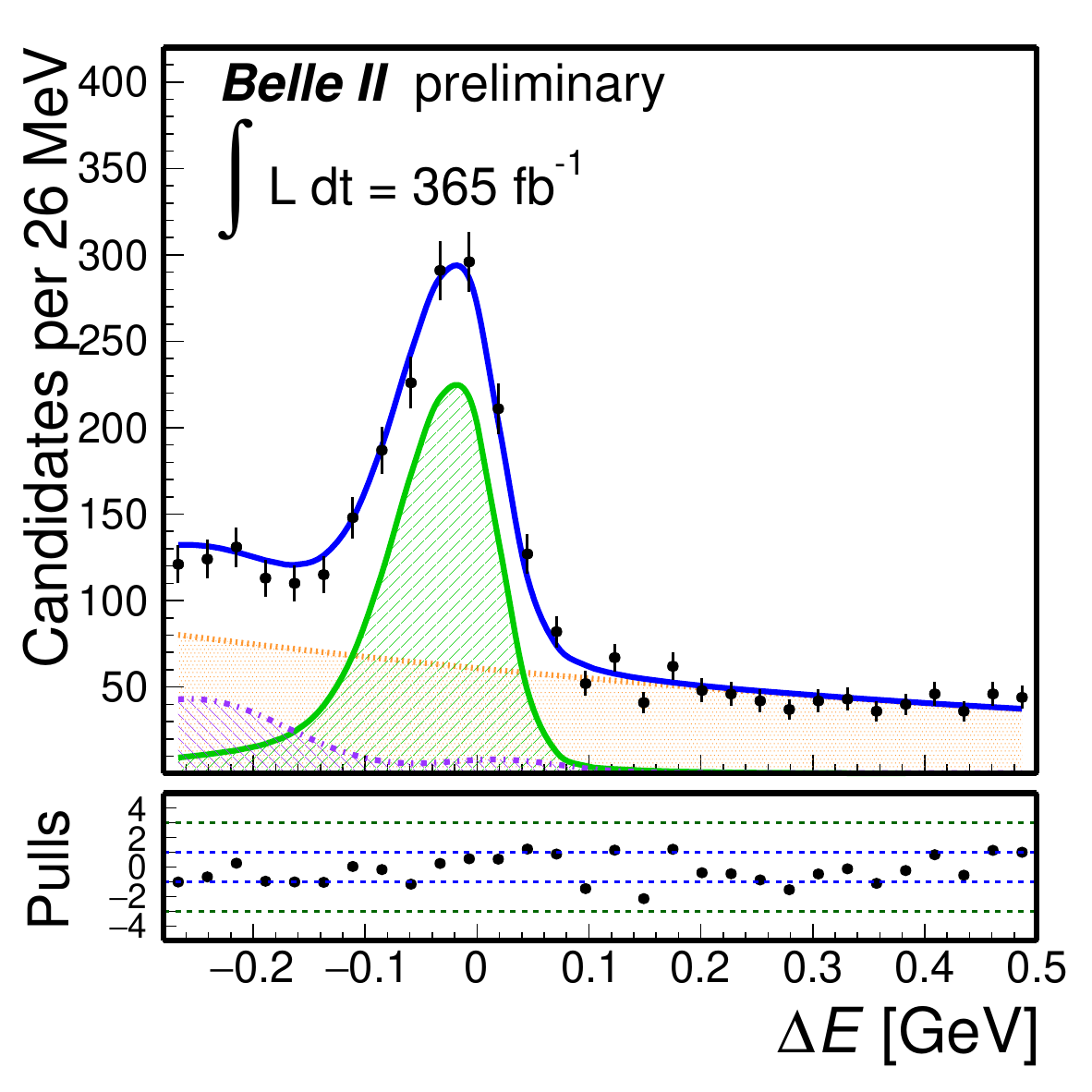}}
   \makebox[0.245\textwidth]{\includegraphics[width=0.245\textwidth]{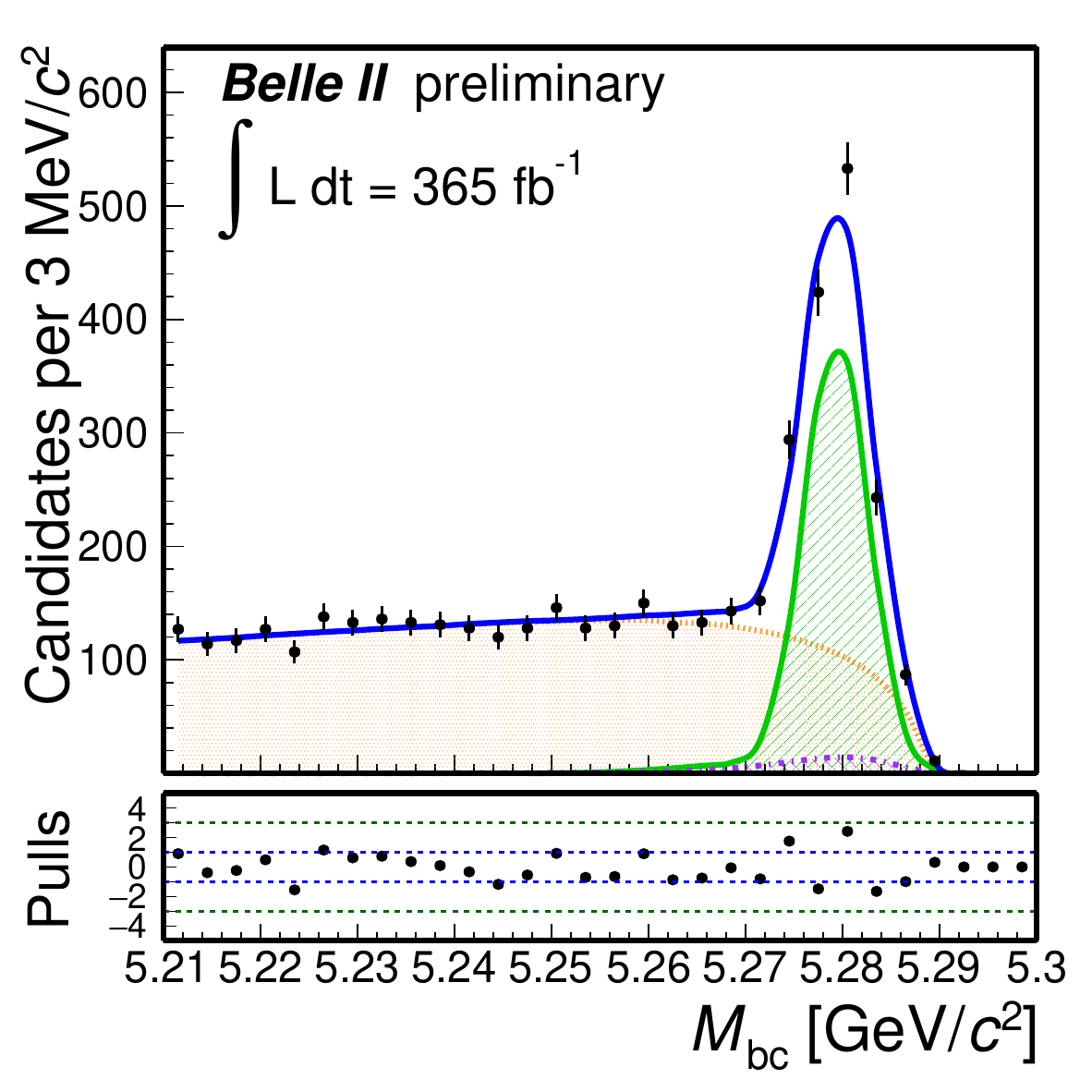}}
   \makebox[0.245\textwidth]{\includegraphics[width=0.245\textwidth]{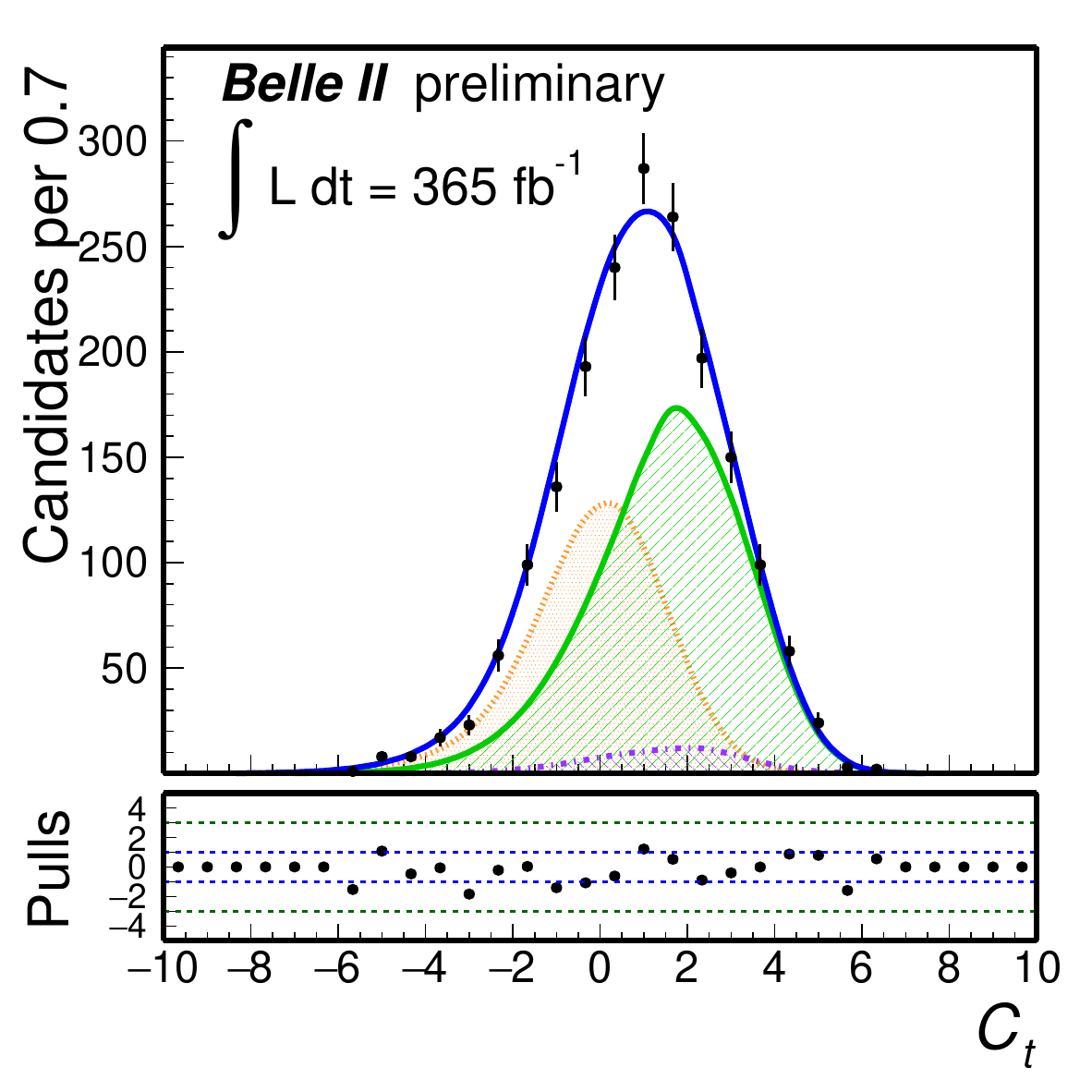}}
   \makebox[0.245\textwidth]{\includegraphics[width=0.245\textwidth]{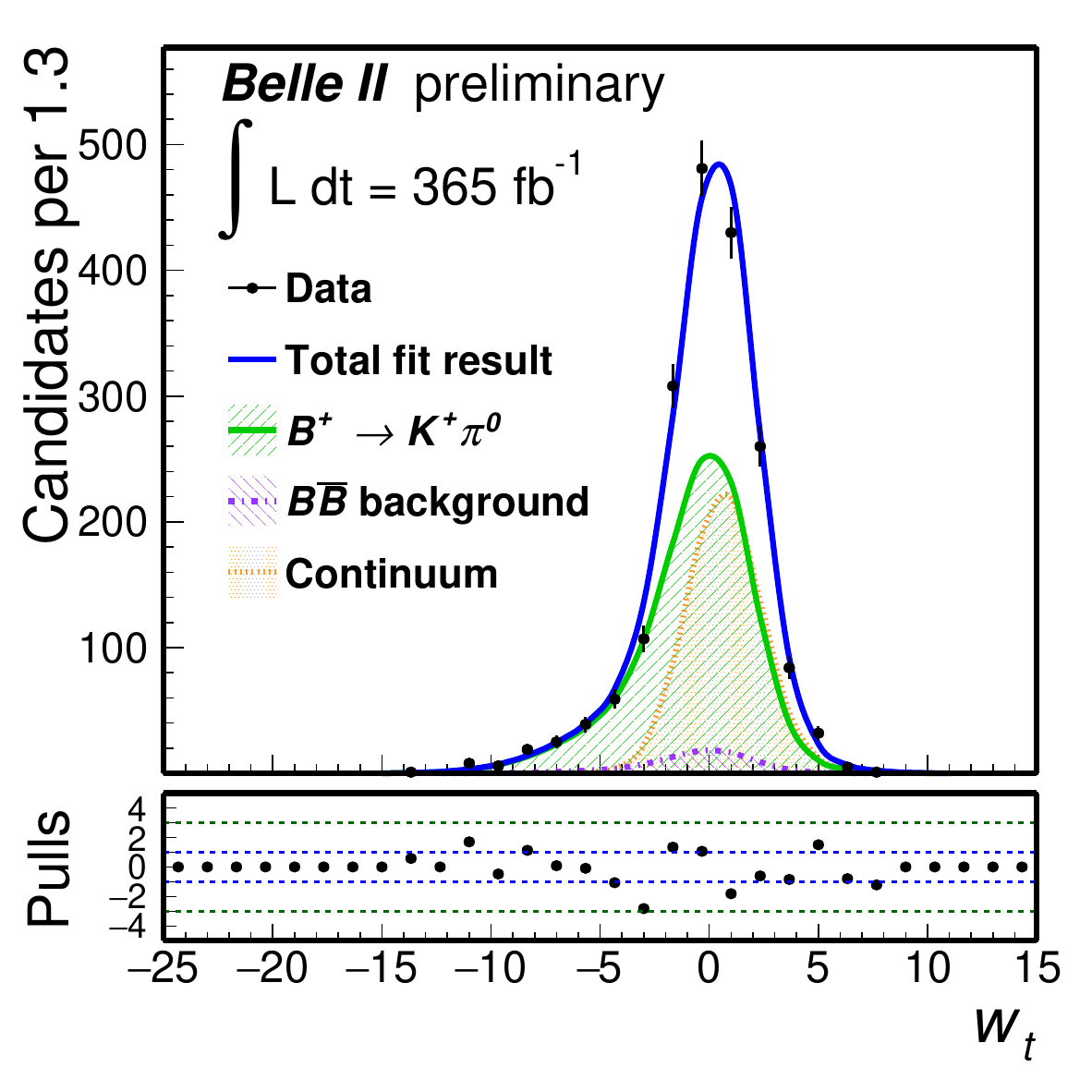}}
    \caption{Signal-enhanced (see text) distributions of (left to right) $\dele$, $\mbc$, $C_{t}$, and $w_t$ for $B^+ \to K^+ \pi^0$ control-sample candidates reconstructed in data, with fit projections overlaid.  Lower panels show differences between observed and best-fit values divided by the fit uncertainties (pulls).
    }
    \label{fig:control_mode}
\end{figure*}
The likelihood function is constructed as
\begin{equation}
\label{eq:pdf}
\begin{aligned}
 \mathcal{L} \propto  \prod_{i=1}^{N}\biggl[&f_{s}\bigl(1 + q^i (1+q^ia_{\rm tag})(1-2\chi_d)(1 -2k(q^i)w^i) 
     {\mathcal{A}}_{\it CP} \bigr) \\
     &\times p_{s}(\Delta E^i,M^i_{\rm bc}) \textcolor{black}{p_{s}(C^i_{t},w^{i}_t)} \\
    & + f_{B\overline{B}}\left(1 + q^i (1+q^ia_{\rm tag})(1 -2k(q^i)w^{i}){\mathcal{A}}_{B\overline{B}}\right) \\
    & \times p_{B\overline{B}}(\Delta E^i)p_{B\overline{B}}(M^i_{\rm bc})\textcolor{black}{p_{B\overline{B}}(C^{i}_{t},w^{i}_t)} \\
    & + (1 - f_{s} - f_{B\overline{B}})\left(1 + q^i(1 -2w^i){\mathcal{A}}_{ c}\right) \\
    & \times \textcolor{black}{p_{c}(\Delta E^i,C^{i}_{t}) p_{c}(M^i_{\rm bc}) p_{c}(w^{i}_t)} \biggr],
\end{aligned}
\end{equation}
where $i$ is the candidate, $N$ is the known total number of candidates in the sample, $f_j$ is the sample fraction for component $j$,  and $p_j(\Delta E^i, M^i_{\rm bc}, C^{i}_{t}, w^i_t, q^i)$ is the probability density function (pdf) for the $i$th candidate to belong to the $j$th component. Here $q^i$ is the predicted flavor of the partner $B$ meson of the $i$th candidate; $w^i$ is the probability for incorrectly identifying the candidate flavor, which dilutes the true asymmetry;  $k(q^i)$ is a multiplicative correction for the flavor-specific difference between predicted and observed $w^i$; and $a_{\rm tag}$ is the $B^0$--$\overline{B}{}^0$ asymmetry in tagging efficiency. The factorization of the pdfs in Eq.~(\ref{eq:pdf}) accounts for observed dependences between observables. The values of $k(q^i)$ and $a_{\rm tag}$ are Gaussian constrained from a flavor-oscillation fit to $B^0 \to D^{(*)-} h^+$ decays in simulation and data, where $h^+$ stands for $\pi^+$ or $K^+$~\cite{GFlaT}. These and all other Gaussian constraints included in our model are omitted from the above likelihood expression to simplify notation. The $\it CP$ asymmetry in data is further diluted by a factor of ($1 - 2\chi_{d}$) due to $\BzBzb$ oscillations. The known time-integrated $\BzBzb$-oscillation probability $\chi_{d} = 0.1858 \pm 0.0011$~\cite{ParticleDataGroup:2024cfk} is Gaussian-constrained in the fit. The effective asymmetry of the $B\overline{B}$ background is constrained to the value ${\mathcal{A}}_{B\overline{B}} = 0.056 \pm 0.095$ as determined in the $\Delta E<-0.3$ GeV and $M_{\rm bc}>5.26$~GeV/$c^2$ sideband. The effective asymmetry of the continuum background ${\mathcal{A}}_{c}$ is freely determined by the fit.

Models for all sample components are chosen based on large samples of simulated events. 
The signal $\mbc$ distribution is modeled using two Gaussian functions. The signal \dele distribution is modeled, in intervals of $M_{\rm bc}$, using the sum of a Gaussian function and a bifurcated Gaussian function. All functions have independent means and widths. The $B\overline{B}$ $\mbc$ distribution is described with the sum of a Gaussian function and a Johnson function~\cite{9d62cdcd}. The continuum $\mbc$ distribution is modeled with the sum of two ARGUS functions~\cite{argus} as the beam energy, which determines the upper
endpoint of the distribution, varies during data taking. The $B\overline{B}$ $\Delta E$ distribution is described with the sum of three Gaussian functions and a bifurcated Gaussian function, while continuum is described with a straight line, with independent slopes in three intervals of $w_t$.  The $C_{t}$ distribution of continuum is modeled using the sum of a Gaussian function and a bifurcated Gaussian function. The  $C_{t}$ distributions of signal and $B\overline{B}$ are each modeled using  the same functional form in three intervals of $w_t$, with independent parameters.
The signal, $B\overline{B}$, and continuum $w_t$ distributions are each modeled using the sum of three Gaussian functions with independent parameters. 
Signal and $B\overline{B}$ model parameters are determined using fits to simulated events. Additional degrees of freedom related to the position and width of the peaking structures are included in the signal-data fit as Gaussian constraints determined from fits to $B^+ \to K^+\pi^0$ decays. These account for residual data-simulation discrepancies such as the $12\pm 4$ MeV difference in \dele peak position.  Continuum model parameters are freely determined by the fit. 

We validate the analysis by applying it to $B^+ \to K^+ \pi^0$ and $B^0 \to \overline{D}{}^0(\to K^+ \pi^- \pi^0) \pi^0$ decays. The photon and $\pi^0$ criteria applied in the $B^+ \to K^+ \pi^0$ selection are the same as for the $B^0 \to \pi^0\pi^0$ analysis. The kaon candidate is a charged particle that satisfies a loose requirement on the ratio $\mathcal{L}_{K}/(\mathcal{L}_{\pi}+\mathcal{L}_{K})$, where the likelihood $\mathcal{L}_{\pi,K}$ for a pion or kaon hypothesis combines particle-identification information from all subdetectors except the pixel detector. For the $B^0 \to \overline{D}{}^0(\to K^+ \pi^- \pi^0) \pi^0$ control channel, all photon and $\pi^0$ selections are the same as for signal except the 1.5 \gevc threshold on $\pi^0$ momentum, which is removed to accommodate the significantly lower momentum spectrum, and an additional $1.84 < m(K^+ \pi^- \pi^0) <$ 1.88 \gevcc restriction on the $D$ mass. Figure~\ref{fig:control_mode} shows signal-enhanced data distributions with fit projections overlaid for the $B^+ \to K^+\pi^0$ channel. The signal-enhancing selection is defined as 5.275 $< \mbc <$ 5.285 \gevcc, $-0.10  < \dele < 0.05\gev$, and $C_{t} > 0$ and applied to all relevant variables except the one displayed. We determine $\mathcal{B}(B^+ \to K^+\pi^0) =(14.3 \pm 0.5)\times 10^{-6}$, \Acp($B^+\to K^+\pi^0) = 0.078 \pm 0.076 $, $\mathcal{B}(B^0 \to \overline{D}{}^0(\to K^+ \pi^- \pi^0) \pi^0) = (41.4 \pm 2.4)\times 10^{-6} $ and $\Acp(B^0 \to \overline{D}{}^0 \pi^0) = 0.01 \pm 0.10$, which all agree with known values~\cite{ParticleDataGroup:2024cfk} within our statistical-only uncertainties. 

\begin{figure*}[th!]
    \makebox[0.245\textwidth]{\includegraphics[width=0.245\textwidth]{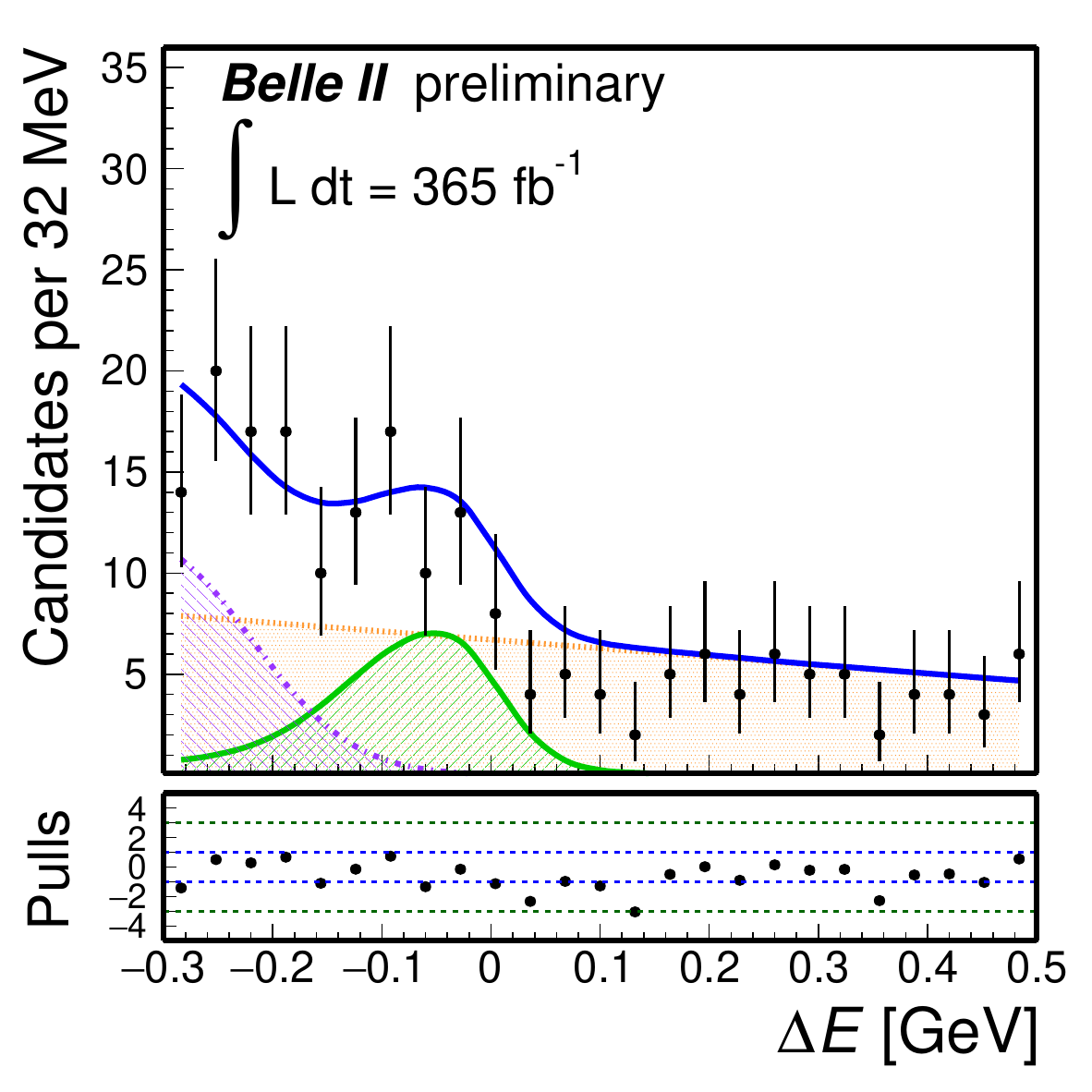}}
    \makebox[0.245\textwidth]
    {\includegraphics[width=0.245\textwidth]{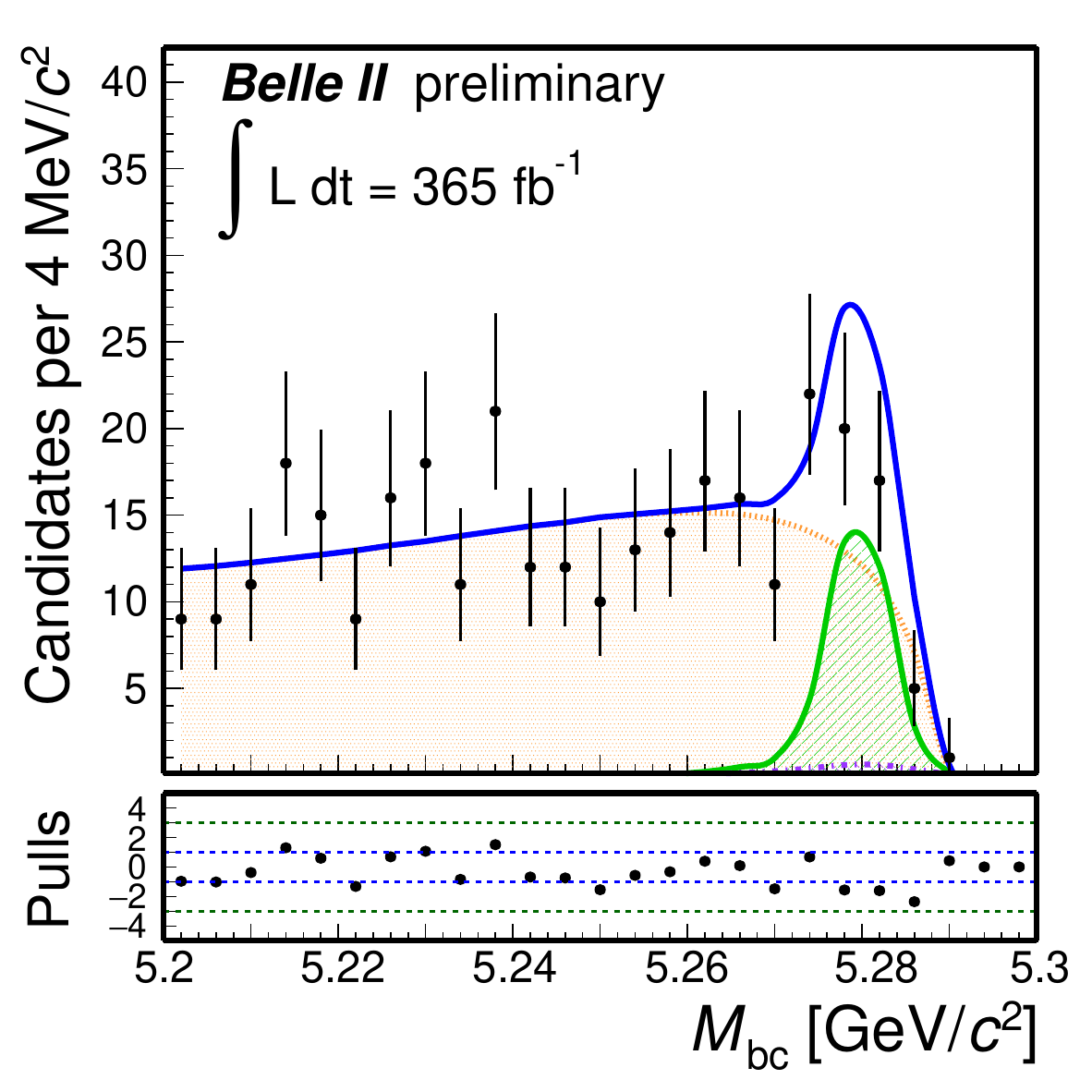}}
    \makebox[0.245\textwidth]{\includegraphics[width=0.245\textwidth]{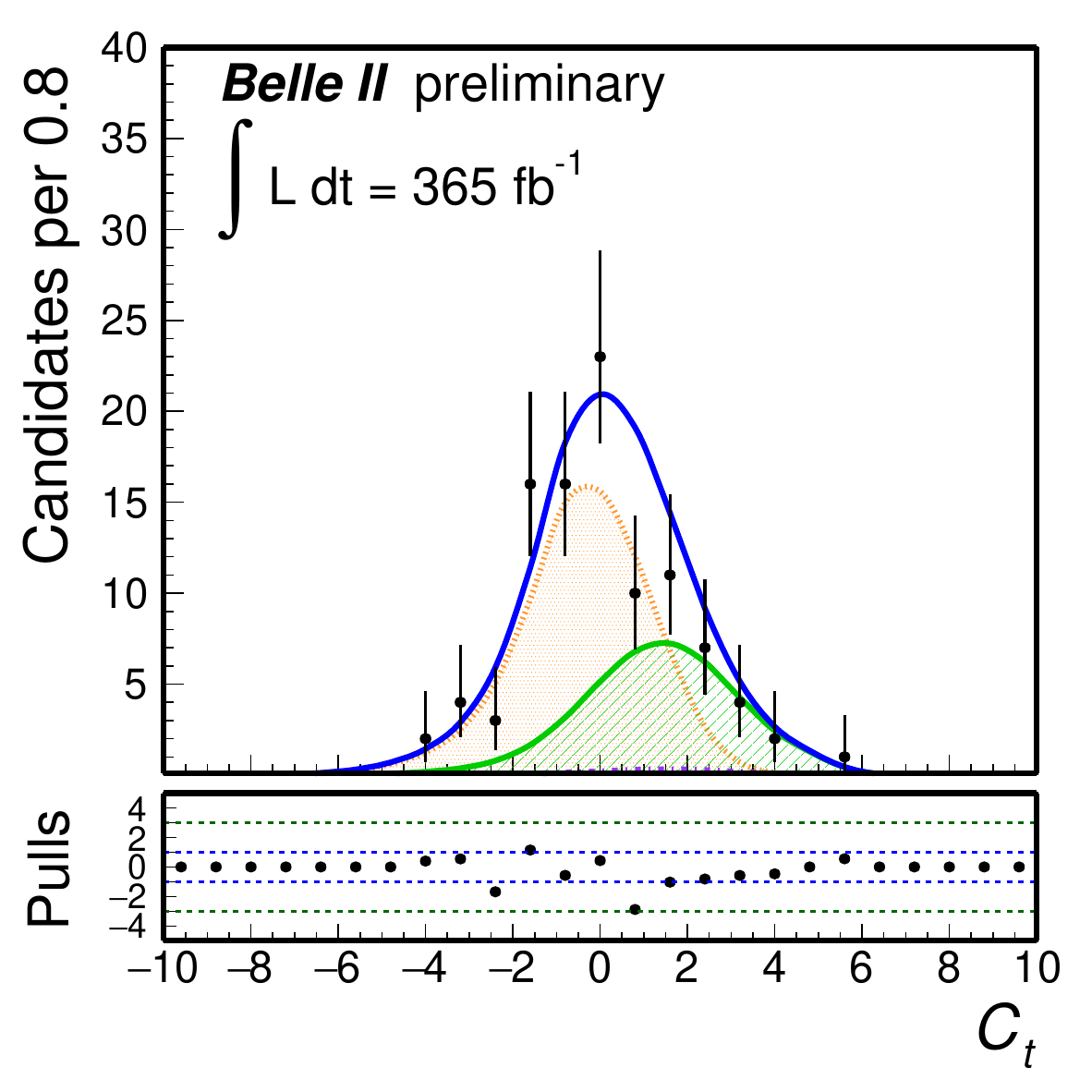}}
    \makebox[0.245\textwidth]{\includegraphics[width=0.245\textwidth]{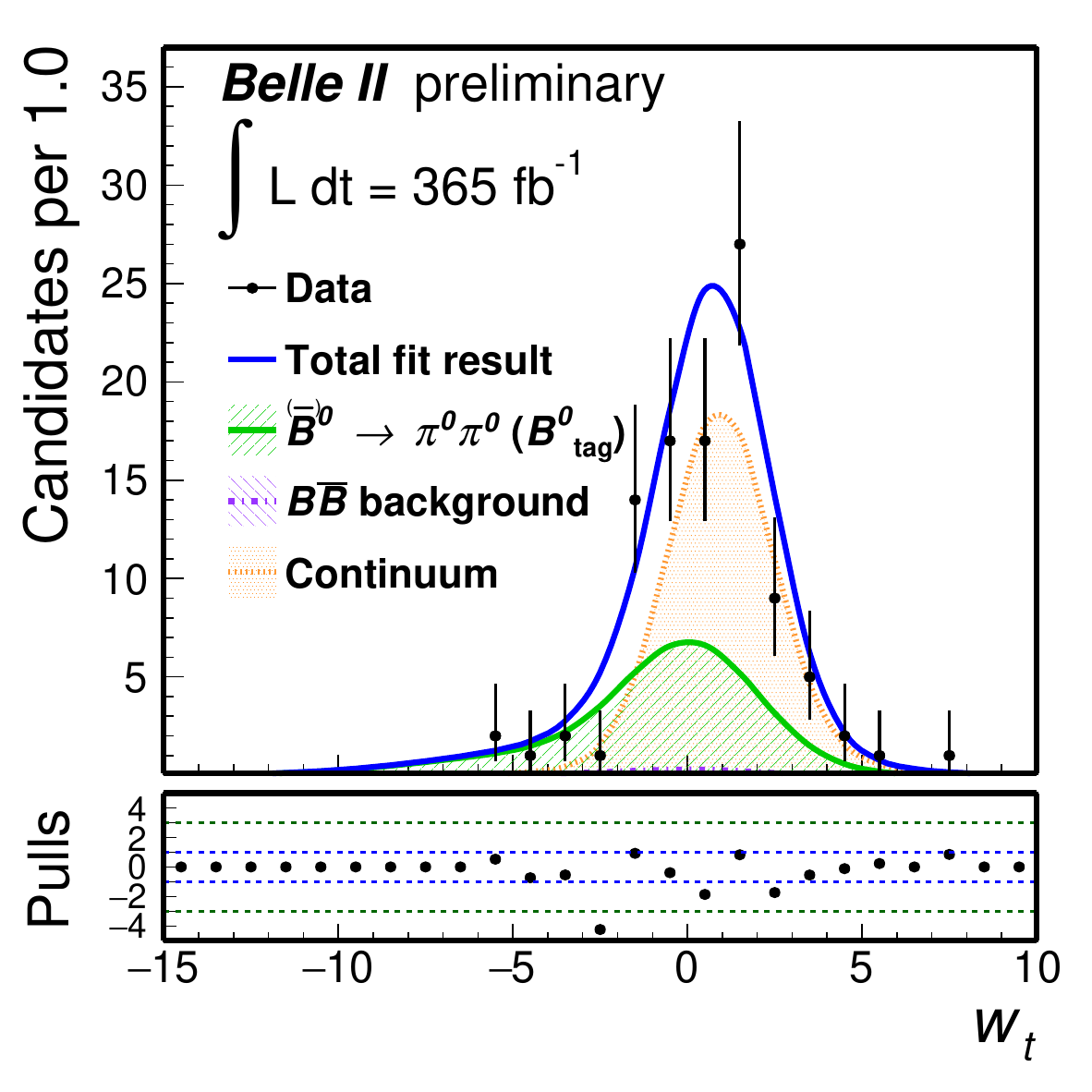}}
    \makebox[0.245\textwidth]{\includegraphics[width=0.245\textwidth]{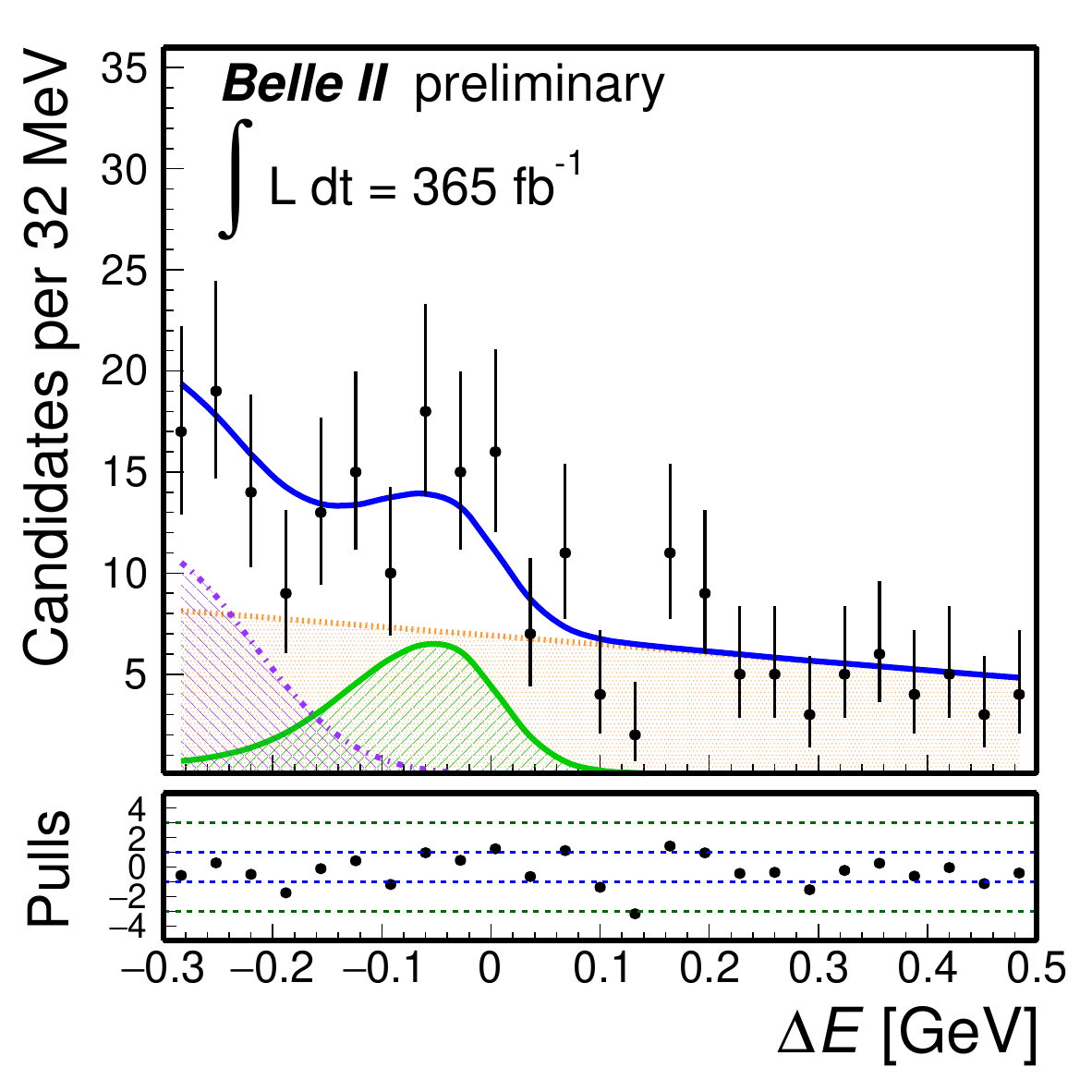}}
    \makebox[0.245\textwidth]
    {\includegraphics[width=0.245\textwidth]{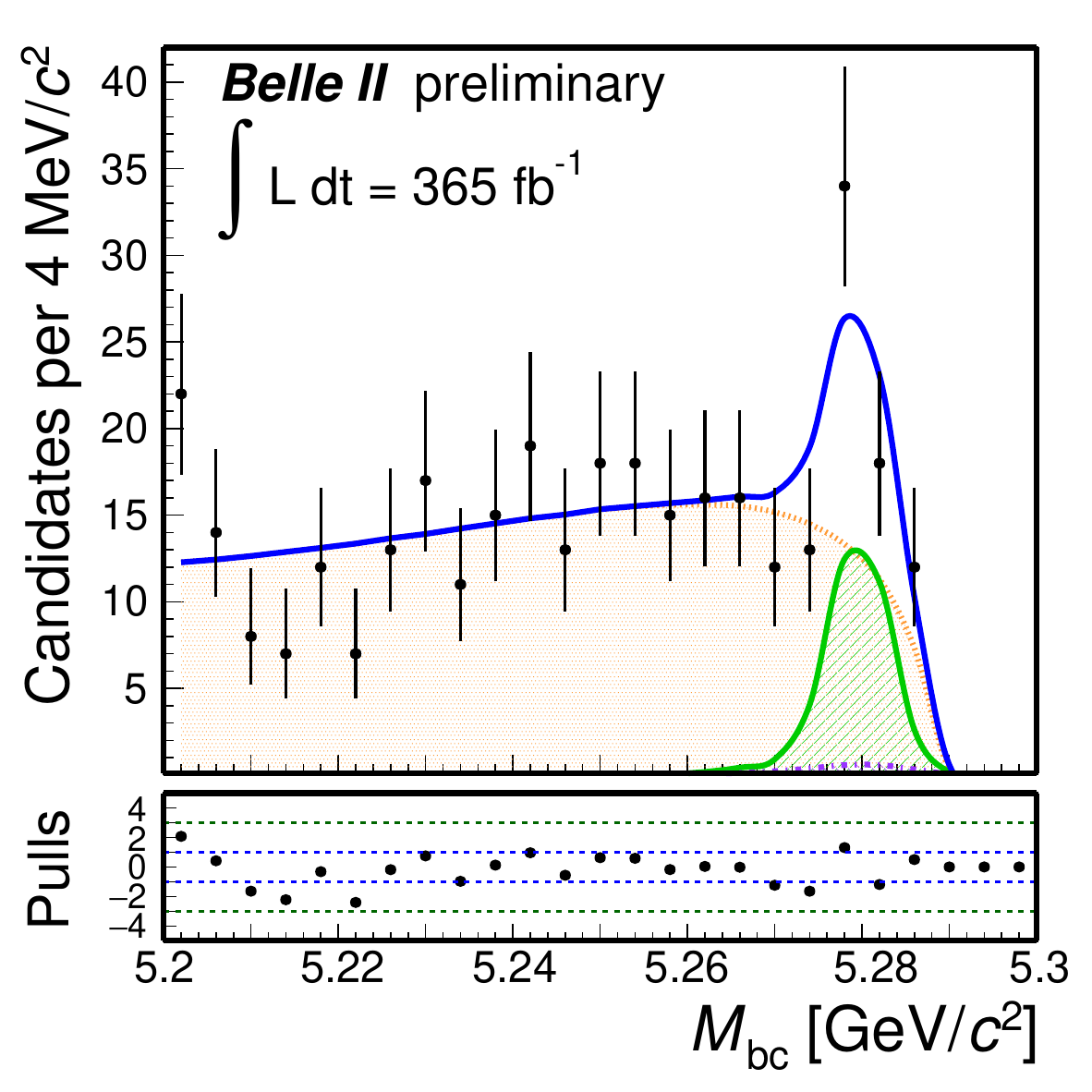}}
    \makebox[0.245\textwidth]{\includegraphics[width=0.245\textwidth]{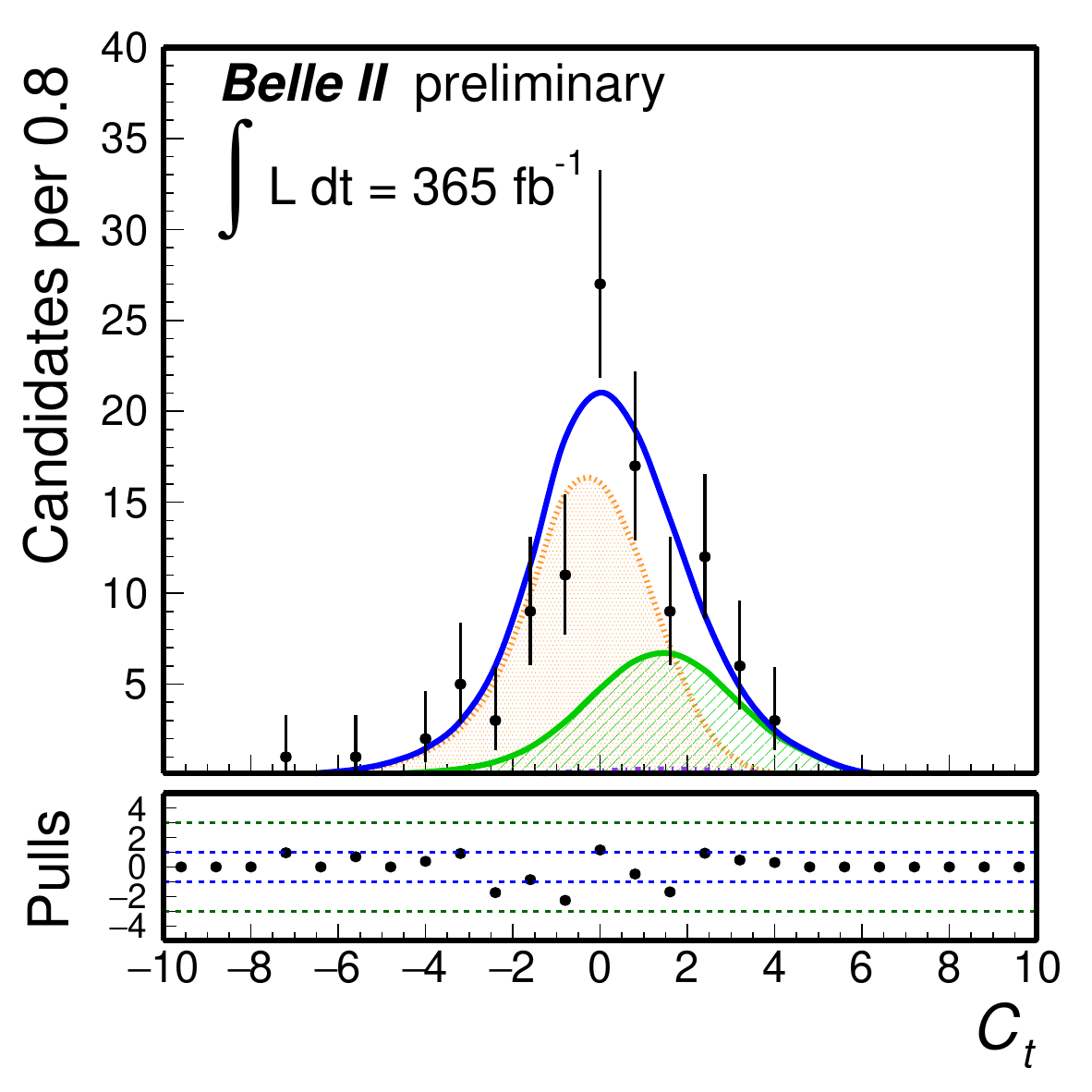}}
    \makebox[0.245\textwidth]{\includegraphics[width=0.245\textwidth]{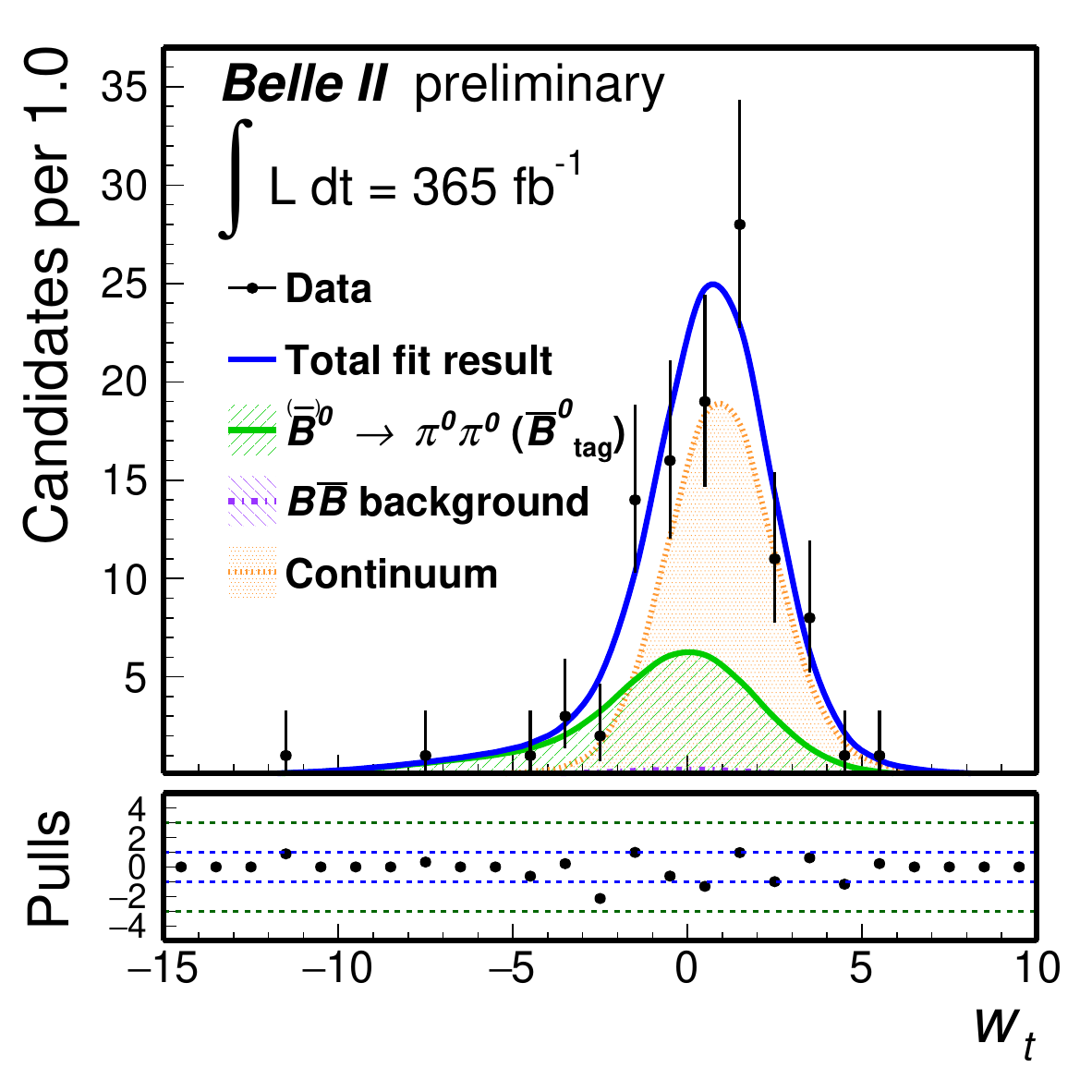}}
    
    \caption{Signal-enhanced distributions of (left to right) $\dele$, $\mbc$, $C_t$, and $w_t$ for  $\Bpipi$ signal candidates with (top) positive and (bottom) negative $q$ tags reconstructed in data, with fit projections overlaid. Lower panels show pulls.}
    \label{fig:ICHEP_fit_split}
\end{figure*}

We maximize the likelihood on the sample of 7140 \Bpipi candidates. The branching fraction, {\it CP} asymmetry, $B\overline{B}$ yield, and continuum yield are freely determined by the fit. Figure~\ref{fig:ICHEP_fit_split} shows signal-enhanced data distributions with fit projections overlaid. The signal-enhancing selection is the same as for the control modes and rejects approximately 91\%--98\% of continuum. A signal is observed in all distributions, overlapping continuum and $B\overline{B}$ background.
The branching fraction is calculated using
\begin{equation}
    \mathcal{B}(\Bpipi) = \frac{N f_{s}(1+f^{+-}/f^{00})}{2 \ \varepsilon \ N_{B\overline{B}} \ \mathcal{B}(\pi^0 \to \gamma \gamma)^2},
\end{equation}
where $N f_{s} = 126 \pm 20$ is the observed signal yield, $\varepsilon = (27.28 \pm 0.03)\%$ is the signal reconstruction and selection efficiency from simulation, $N_{B\overline{B}} = (387 \pm 6) \times 10^{6}$ is the number of $B\overline{B}$ pairs in the sample, $f^{+-}/f^{00} = 1.052 \pm 0.031$ is the ratio of the branching fractions for the decay of $\Y4S$ to $\BpBm$ and $\Bz \overline{B}{}^0$ final states ~\cite{banerjee2024averagesbhadronchadrontaulepton2}, and $\mathcal{B}(\pi^0 \to \gamma \gamma) = (98.82 \pm 0.03)\%$~\cite{ParticleDataGroup:2024cfk} is the relevant $\pi^0$ branching fraction. The Poisson fluctuation on $N$ is included in the branching-fraction statistical uncertainty to account for fluctuations of the total sample size. We obtain $\mathcal{B}(\Bpipi) = (1.25 \pm 0.20)\times 10^{-6}$ and $\Acp(\Bpipi) = 0.03 \pm 0.30$
where uncertainties are statistical. 

We consider sources of systematic uncertainties associated with assumptions made in the analysis, with possible biases due to discrepancies between relevant distributions in data and simulation, or with intrinsic uncertainties of external inputs (Table~\ref{table:uncertainty}). Whenever a systematic source is associated with a modeling choice in the analysis, we determine its impact by using ensembles of simplified simulated experiments. 
To account  for possible data-simulation discrepancies, we use control samples reconstructed in data and in simulation to estimate correction factors and assess their associated uncertainties, which are propagated in the results as systematic uncertainties. We also propagate uncertainties on the external inputs to the quantities of interest.  

A systematic uncertainty of 8.1\% associated with the $\piz$ pair reconstruction efficiency is determined from data using the decays $D^{*-} \to \overline{D}{}^0(\to K^+ \pi^- \pi^0) \pi^-$ and $D^{*-} \to \overline{D}{}^0(\to K^+ \pi^-) \pi^-$. 
The correction factor that matches the efficiency of the continuum-classifier selection in simulation with data is determined using $B^+ \to \overline{D}{}^0(\to K^+ \pi^-) \pi^+$ decays. The uncertainty on the correction factor is assigned as a systematic uncertainty. The systematic uncertainty associated with modeling choices, and with the uncertainties on the associated pdf parameters, is the maximum difference observed between averages of results obtained from fitting our model to simulated data generated with alternative functions and the average obtained with the nominal parameter and model choices, or with parameters varied according to their covariances. The continuum component has no contribution from alternate parameter choices, as its parameters are directly determined in the fit.  The systematic uncertainty on $f^{+-}/f^{00}$ accounts for the experimental uncertainty on the measurement and the uncertainty due to the assumption of isospin symmetry. The uncertainty in the number of $B\overline{B}$ meson pairs is due to the uncertainty in the integrated-luminosity determination, the small amount of off-resonance data, and efficiency mismodeling in simulation. The uncertainty on the time-integrated $\BzBzb$-oscillation probability $\chi_{d}$ is assessed by repeating the fit after fixing the parameter to its best-fit value, and subtracting in quadrature the uncertainties on the parameters of interest. 
The systematic uncertainty associated with the choice of calibration function used to obtain the factors $k(q)$ is determined by fitting simulated data generated with an alternative function, which includes additional corrections proportional to $w^2$.
The total systematic uncertainty is the sum in quadrature of the individual contributions.

\begin{table}[h]
\centering
\caption{Fractional systematic uncertainties on the branching fraction and absolute systematic uncertainties on the {\it CP} asymmetry. Total systematic uncertainties, resulting from their sums in quadrature, are also given and compared with statistical uncertainties.}
\begin{tabular}{l c c c}
\hline\hline
 Source & $\mathcal{B}$ &  $\mathcal{A}_{\it CP}$\\
\hline
$\pi^0$ efficiency      & 8.1\%  & n/a\\
Continuum-suppression efficiency & 1.9\% & n/a\\
$B\overline{B}$-background model &  1.7\%   & 0.01\\
Signal model & 1.2\%   & 0.02\\
Continuum-background model &  0.9\%   & 0.03\\
\Y4S branching fractions $(1+f^{+-}/f^{00})$      &  1.5 \% &  n/a\\
Sample size $N_{B\bar{B}}$      &  1.5\%  &  n/a\\
$\BzBzb$-oscillation probability      &  n/a  & $<0.01$ \\ 
Wrong-tag probability calibration & n/a & 0.01\\
\hline
Total systematic uncertainty & 8.9\%   & 0.04 \\
\hline
Statistical uncertainty & 15.9\% & 0.30  \\
\hline\hline
\end{tabular} 
\label{table:uncertainty}
\end{table}

The final results are 
\begin{equation}
\mathcal{B}(\Bpipi) = (1.25 \pm 0.20 \pm 0.11)\times 10^{-6}
\end{equation} and 
\begin{equation}
\Acp(\Bpipi) = 0.03 \pm 0.30 \pm 0.04,
\end{equation} 
where the first contributions to the uncertainties are statistical and the second systematic. The (statistical) linear correlation between these two quantities is $+1.5\%$. This work supersedes a previous Belle II result based on about one-half of the data sample~\cite{PhysRevD.107.112009}, and incorporates a number of enhancements.  We improve background suppression by increasing the discriminating power of classifiers; we simplify the sample-composition fit by including as an observable the predicted signal flavor obtained by new algorithms with higher efficacy; we include in the fit additional control data to constrain backgrounds directly from data, thus reducing systematic uncertainties. These improvements reduce the fractional systematic uncertainty on the branching fraction by 40\% and both the absolute statistical and systematic uncertainty on the {\it CP} asymmetry by 3\%, for a given sample size.  Combining these improvements with the increased sample size produces results competitive with the current best values, which are based on larger samples.
We average our results with previous measurements of $B^0 \to \pi^0 \pi^0$  branching fraction and {\it CP} asymmetry apart
from the previous Belle II results~\cite{PhysRevD.107.112009},  and include them, along with recent $B^0 \to \pi^+ \pi^-$ and $B^+ \to \pi^+ \pi^0$  inputs~\cite{ParticleDataGroup:2024cfk,PhysRevD.109.012001}, in an isospin analysis that follows Ref.~\cite{Gronau1990} to assess impact on $B \to \pi\pi$-based $\phi_2$ constraints.  Our results reduce by 10$^\circ$ the  68\% CL exclusion interval surrounding the CKM-favored solution, corresponding to a 30\% fractional increase in $\phi_2$ precision~\footnote{See Supplemental Material for the impact on $\phi_2$ determination, which includes Ref.~\cite{PhysRevD.107.052008}}. This makes the precision of $\phi_2$ determinations based on $B \to \pi\pi$ decays competitive with the precision of $B \to \rho \rho$ determinations, resulting in a global improvement on the $\phi_2$ precision.

In summary, we report an improved Belle II measurement of the branching fraction and direct $\it CP$ asymmetry of $B^0 \to \pi^0\pi^0$ decays reconstructed in the full electron-positron collision sample at the \Y4S collected through 2022. The results are $\mathcal{B}(\Bpipi) = (1.25 \pm 0.23)\times 10^{-6}$ and  $\Acp(\Bpipi) = 0.03 \pm 0.30$. These measurements achieve a precision superior to, or comparable with, the precision of previous measurements, based on larger samples, and advance our knowledge of two-body charmless $B$ decays and of the angle $\phi_2$.

This work, based on data collected using the Belle II detector, which was built and commissioned prior to March 2019,
was supported by
Higher Education and Science Committee of the Republic of Armenia Grant No.~23LCG-1C011;
Australian Research Council and Research Grants
No.~DP200101792, 
No.~DP210101900, 
No.~DP210102831, 
No.~DE220100462, 
No.~LE210100098, 
and
No.~LE230100085; 
Austrian Federal Ministry of Education, Science and Research,
Austrian Science Fund
No.~P~34529,
No.~J~4731,
No.~J~4625,
and
No.~M~3153,
and
Horizon 2020 ERC Starting Grant No.~947006 ``InterLeptons'';
Natural Sciences and Engineering Research Council of Canada, Compute Canada and CANARIE;
National Key R\&D Program of China under Contract No.~2022YFA1601903,
National Natural Science Foundation of China and Research Grants
No.~11575017,
No.~11761141009,
No.~11705209,
No.~11975076,
No.~12135005,
No.~12150004,
No.~12161141008,
No.~12475093,
and
No.~12175041,
and Shandong Provincial Natural Science Foundation Project~ZR2022JQ02;
the Czech Science Foundation Grant No.~22-18469S 
and
Charles University Grant Agency project No.~246122;
European Research Council, Seventh Framework PIEF-GA-2013-622527,
Horizon 2020 ERC-Advanced Grants No.~267104 and No.~884719,
Horizon 2020 ERC-Consolidator Grant No.~819127,
Horizon 2020 Marie Sklodowska-Curie Grant Agreement No.~700525 ``NIOBE''
and
No.~101026516,
and
Horizon 2020 Marie Sklodowska-Curie RISE project JENNIFER2 Grant Agreement No.~822070 (European grants);
L'Institut National de Physique Nucl\'{e}aire et de Physique des Particules (IN2P3) du CNRS
and
L'Agence Nationale de la Recherche (ANR) under grant ANR-21-CE31-0009 (France);
BMBF, DFG, HGF, MPG, and AvH Foundation (Germany);
Department of Atomic Energy under Project Identification No.~RTI 4002,
Department of Science and Technology,
and
UPES SEED funding programs
No.~UPES/R\&D-SEED-INFRA/17052023/01 and
No.~UPES/R\&D-SOE/20062022/06 (India);
Israel Science Foundation Grant No.~2476/17,
U.S.-Israel Binational Science Foundation Grant No.~2016113, and
Israel Ministry of Science Grant No.~3-16543;
Istituto Nazionale di Fisica Nucleare and the Research Grants BELLE2,
and
the ICSC – Centro Nazionale di Ricerca in High Performance Computing, Big Data and Quantum Computing, funded by European Union – NextGenerationEU;
Japan Society for the Promotion of Science, Grant-in-Aid for Scientific Research Grants
No.~16H03968,
No.~16H03993,
No.~16H06492,
No.~16K05323,
No.~17H01133,
No.~17H05405,
No.~18K03621,
No.~18H03710,
No.~18H05226,
No.~19H00682, 
No.~20H05850,
No.~20H05858,
No.~22H00144,
No.~22K14056,
No.~22K21347,
No.~23H05433,
No.~26220706,
and
No.~26400255,
and
the Ministry of Education, Culture, Sports, Science, and Technology (MEXT) of Japan;  
National Research Foundation (NRF) of Korea Grants
No.~2016R1-D1A1B-02012900,
No.~2018R1-A6A1A-06024970,
No.~2021R1-A6A1A-03043957,
No.~2021R1-F1A-1060423,
No.~2021R1-F1A-1064008,
No.~2022R1-A2C-1003993,
No.~2022R1-A2C-1092335,
No.~RS-2023-00208693,
No.~RS-2024-00354342
and
No.~RS-2022-00197659,
Radiation Science Research Institute,
Foreign Large-Size Research Facility Application Supporting project,
the Global Science Experimental Data Hub Center, the Korea Institute of
Science and Technology Information (K24L2M1C4)
and
KREONET/GLORIAD;
Universiti Malaya RU grant, Akademi Sains Malaysia, and Ministry of Education Malaysia;
Frontiers of Science Program Contracts
No.~FOINS-296,
No.~CB-221329,
No.~CB-236394,
No.~CB-254409,
and
No.~CB-180023, and SEP-CINVESTAV Research Grant No.~237 (Mexico);
the Polish Ministry of Science and Higher Education and the National Science Center;
the Ministry of Science and Higher Education of the Russian Federation
and
the HSE University Basic Research Program, Moscow;
University of Tabuk Research Grants
No.~S-0256-1438 and No.~S-0280-1439 (Saudi Arabia), and
King Saud University,Riyadh, Researchers Supporting Project number (RSPD2024R873)  
(Saudi Arabia);
Slovenian Research Agency and Research Grants
No.~J1-9124
and
No.~P1-0135;
Agencia Estatal de Investigacion, Spain
Grant No.~RYC2020-029875-I
and
Generalitat Valenciana, Spain
Grant No.~CIDEGENT/2018/020;
The Knut and Alice Wallenberg Foundation (Sweden), Contracts No.~2021.0174 and No.~2021.0299;
National Science and Technology Council,
and
Ministry of Education (Taiwan);
Thailand Center of Excellence in Physics;
TUBITAK ULAKBIM (Turkey);
National Research Foundation of Ukraine, Project No.~2020.02/0257,
and
Ministry of Education and Science of Ukraine;
the U.S. National Science Foundation and Research Grants
No.~PHY-1913789 
and
No.~PHY-2111604, 
and the U.S. Department of Energy and Research Awards
No.~DE-AC06-76RLO1830, 
No.~DE-SC0007983, 
No.~DE-SC0009824, 
No.~DE-SC0009973, 
No.~DE-SC0010007, 
No.~DE-SC0010073, 
No.~DE-SC0010118, 
No.~DE-SC0010504, 
No.~DE-SC0011784, 
No.~DE-SC0012704, 
No.~DE-SC0019230, 
No.~DE-SC0021274, 
No.~DE-SC0021616, 
No.~DE-SC0022350, 
No.~DE-SC0023470; 
and
the Vietnam Academy of Science and Technology (VAST) under Grants
No.~NVCC.05.12/22-23
and
No.~DL0000.02/24-25.

These acknowledgements are not to be interpreted as an endorsement of any statement made
by any of our institutes, funding agencies, governments, or their representatives.

We thank the SuperKEKB team for delivering high-luminosity collisions;
the KEK cryogenics group for the efficient operation of the detector solenoid magnet and IBBelle on site;
the KEK Computer Research Center for on-site computing support; the NII for SINET6 network support;
and the raw-data centers hosted by BNL, DESY, GridKa, IN2P3, INFN, 
and the University of Victoria.

\bibliographystyle{apsrev4-1}
\bibliography{belle2}

\section*{Supplemental Material}
\subsection*{Impact on $\phi_2$ determination}

\begin{figure}[b!]
    \makebox[0.49\textwidth]{\includegraphics[width=0.49\textwidth]{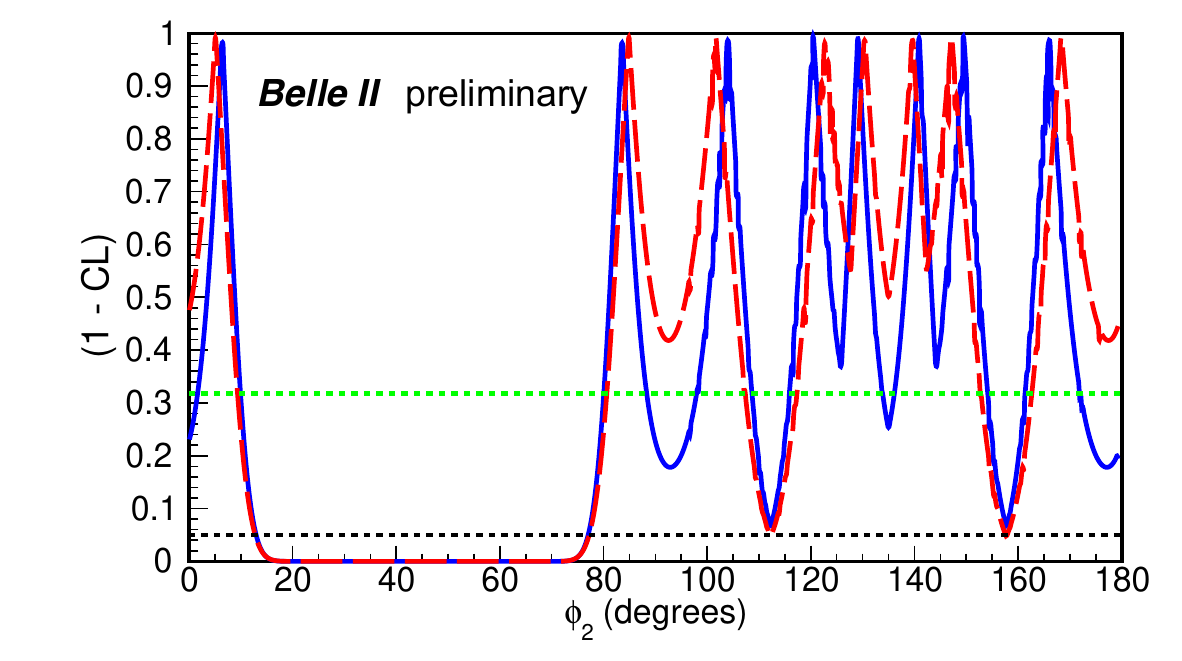}}
    \caption{P-value as a function of the CKM angle $\phi_2$ from an isospin-based combination of all $B \to \pi\pi$ results (red dashed) without and (blue solid) with the inclusion of the results of this work.}
    \label{fig:alpha_pipi}
\end{figure}

We assess the impact of the results in terms of changes in the constraints on the CKM angle $\phi_2$ determined solely from $B \to \pi\pi$ isospin relations.
Figure~\ref{fig:alpha_pipi} compares the p-value as a function of $\phi_2$ resulting from the combination of all $B \to \pi\pi$ results available until 2023~\cite{PhysRevD.107.052008} with the 2024 combination in which the existing Belle~II $B^0 \to \pi^0 \pi^0$ measurements~\cite{PhysRevD.107.112009} are replaced with the results of this work. 

\end{document}